\documentclass{iopart}

\expandafter\let\csname equation*\endcsname\relax
\expandafter\let\csname endequation*\endcsname\relax

\usepackage{lipsum} 
\usepackage{color}
\usepackage{graphicx}
\usepackage{float}
\usepackage{iopams}
\usepackage[margin=10pt,font=small,format=hang]{caption}
\usepackage{amsmath}
\usepackage{acronym}
\usepackage{multirow}
\usepackage{tabu}
\usepackage{tikz}
\usetikzlibrary{arrows,shapes,trees,decorations.pathreplacing}
\usepackage{import}
\usepackage{svn-multi}
\usepackage{lineno}
\usepackage{hyperref}
\usepackage{mathtools}
\usepackage{url}
\hypersetup{
    colorlinks=true,
    linkcolor=blue,
    filecolor=magenta,      
    urlcolor=cyan,
}

\DeclareGraphicsExtensions{%
    .pdf,.PDF,%
    .png,.PNG,%
    .jpg,.mps,.jpeg,.jbig2,.jb2,.JPG,.JPEG,.JBIG2,.JB2}

\begin{document}

\title[]{Improving the Sensitivity of Advanced LIGO Using Noise Subtraction}
\author{Derek Davis$^1$,
        Thomas Massinger$^2$,
        Andrew Lundgren$^3$,
        Jennifer C. Driggers$^2$,
        Alex L. Urban$^4$,
        Laura Nuttall$^{3,5}$
        }

\address{$^1$Syracuse University, Syracuse, NY 13244, USA}
\address{$^2$California Institute of Technology, Pasadena, CA 91125, USA}
\address{$^3$University of Portsmouth, Portsmouth PO1 2UP, United Kingdom}
\address{$^4$Louisiana State University, Baton Rouge, LA 70803, USA}
\address{$^5$Cardiff University, Cardiff CF24 3AA, United Kingdom}
\date{\today}

\begin{abstract}
This paper presents an adaptable, parallelizable method 
for subtracting linearly coupled noise from Advanced LIGO data.
We explain the features developed to ensure that the process is robust 
enough to handle the variability present in Advanced LIGO data. In this work, 
we target subtraction of noise due to beam jitter, detector calibration lines, 
and mains power lines. We demonstrate 
noise subtraction over the entirety of the second observing run, resulting in 
increases in sensitivity comparable to those reported in previous targeted efforts. 
Over the course of the second observing run, we see a 30\% increase in 
Advanced LIGO sensitivity to gravitational waves from a broad 
range of compact binary systems.
We expect the use of this method to result in a higher 
rate of detected gravitational-wave signals in Advanced LIGO data. 
\end{abstract}

\maketitle

\section{Introduction}\label{s:intro}

Advanced LIGO's (aLIGO) second observing run (O2) lasted from November 30, 2016 to August 26, 2017. 
Initial analysis of the O2 data set resulted in the detection of gravitational-wave 
signals from 3 binary black hole (BBH) systems \cite{GW170104, GW170608, GW170814} and the 
first ever detection of gravitational waves from a binary neutron star (BNS) system \cite{GW170817}.

It has been 
previously shown that it is possible to increase the sensitivity of the aLIGO detectors by 
subtracting instrumental noise from the gravitational-wave strain data 
\cite{Driggers:2012noise,DeRosa:2012,Tiwari:2015,Meadors:2014,Coughlin:2016}. 
For source parameter estimation \cite{Veitch:2015lal} of previously published gravitational-wave signals from O2, a MATLAB-based noise 
subtraction algorithm was used to subtract instrumental noise using an associated witness sensor 
for 4096 seconds around identified events \cite{Driggers:2017,Driggers:2018pe}. This was the first 
instance of noise subtraction being used in the analysis of gravitational-wave events. However, this 
process was not designed with the intention of subtracting noise from the entire O2 data set.

Since this initial analysis, a Python-based implementation of noise subtraction was developed 
that prioritizes parallel processing and computational efficiency with the goal of subtracting 
instrumental noise over the entirety of the second observing run. Considering that each 
individual interferometer recorded over 150 days of data, one of the key considerations was 
the size of the data set that this noise subtraction pipeline needed to process. 
The methods used in this pipeline are general enough to allow any linearly coupled noise source 
with a clear witness to be subtracted out efficiently. 

This manuscript describes the method used to subtract noise due to beam jitter, detector calibration 
lines, and mains power lines in O2 and reports the improvement to search sensitivity gained by applying 
this method. 
Section \ref{sec:methodology} outlines the workflow used to process 
the data set in parallel. Section \ref{sec:noise} characterizes the instrumental noise sources that 
were subtracted from the O2 data set. Section \ref{sec:diagnostics} describes the tests that 
were done to ensure that the subtraction process was not capable of removing genuine astrophysical 
signals. Section \ref{sec:results} presents the effects of noise 
subtraction on the aLIGO noise spectrum and on the sensitivity to simulated 
astrophysical signals.

\section{Subtraction Pipeline Overview}\label{sec:methodology}

\subsection{Measurement of Transfer Functions}

\begin{figure}[t]
\centering
\includegraphics[width=\textwidth]{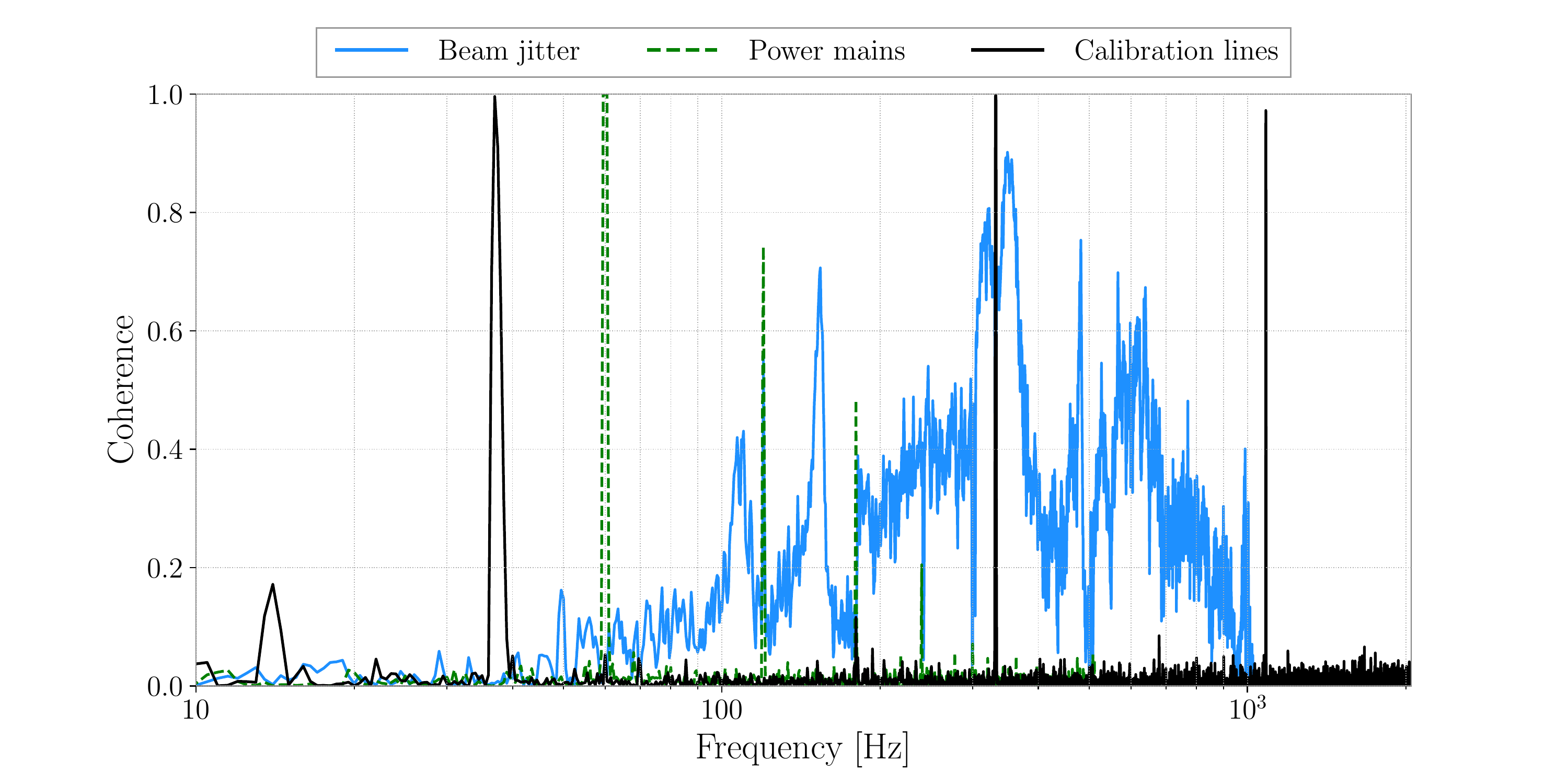}
\caption{Coherence between witness sensors and gravitational wave strain for three %
types of instrumental noise subtracted in O2 at Hanford: beam jitter, power mains, and %
calibration lines. The measured coherence demonstrates significant linear %
coupling between these witness sensors and the strain data, motivating the %
use of linear subtraction methods.}
\label{fig:coherence}
\end{figure}

The assumption of a linear transfer function is motivated by the high coherence 
between witness sensor signals and gravitational-wave strain data. 
Figure \ref{fig:coherence} shows the coherence between witness sensors and 
gravitational-wave strain for three types of instrumental noise subtracted in 
O2. These noise sources are further detailed in Section \ref{sec:noise}. 

For a given noise source, we assume that our measured gravitational wave strain 
data, $h(t)$, contains a noise component 
that can be modeled as the convolution of an unknown transfer function $c'(t)$ and 
the output of a witness sensor $a(t)$, 

\begin{equation}
h(t) = h'(t) + a(t) * c'(t).
\end{equation}
This noise component can be 
removed from the strain data by filtering the witness 
sensor data with this transfer function and subtracting its contribution to 
the measured strain, resulting in a residual strain denoted $h'(t)$. This transfer function 
can be conveniently calculated in the frequency domain, so that the subtraction takes the form 

\begin{equation}\label{eq:freqsub}
\tilde{h}(f) = \tilde{h}'(f) + \tilde{a}(f) \cdot \tilde{c}'(f).
\end{equation}

Adapting the methodology and notation from \cite{Allen:1999ct}, we begin by considering our 
data as time series that are sampled 
at time interval $\Delta t$ over a time period $T$. This results in $M=T/\Delta t$ samples, 
denoted by $Y(j)$ for $j=0, ..., M-1$. We denote the Discrete Fourier Transforms of each 
data stream as  $\tilde{Y}(k)$ for $k=-M/2,...,M/2$, so that the $k$'th bin corresponds 
to a frequency $f = k/T$. We then split the frequency space into bands of 
width $F$ given by 

\begin{equation}
f \in [f_b,f_{b+1}) \text{ with } f_b = \frac{bF}{T}
\end{equation}
for $b = 0,...,M/2F$. The transfer function is measured independently over each of 
these frequency bands and is constructed using frequency domain inner products between 
the relevant data sets. 
For two data sets $Y_1$ and $Y_2$ the inner product over 
a specific frequency band $b$ is calculated as the cross-power spectrum summed over 
that frequency band:

\begin{equation}
\tilde{c}_{12}(f_b) = \sum_{f = f_b}^{f_{(b+1)}} \tilde{Y_1}(f) \tilde{Y_2}^*(f).
\end{equation}

A measurement of the transfer function for uncorrelated noise, for which 
each frequency bin has a random phase, should find no significant coupling 
as multiple uncorrelated data points are averaged over to calculate the transfer function. 
To help reduce the risk of spurious 
correlations being measured, we set a minimum threshold on the value that the transfer 
function can take as a fraction of the maximum value and set the value of the transfer 
function to zero in any band whose value is below that threshold.
We found that a uniform fractional threshold of 
$2.5 \times 10^{-9}$ was sufficient for the noise sources considered in this work, 
but in practice this value can be tuned
for different use cases. 

In the case when multiple sensors witness the same noise, there will be a measurable 
correlation between each of the sensors, resulting in oversubtraction if not accounted 
for. For $N$ witness sensors $Y_1,...Y_N$ and a target data stream to subtract noise from, 
$Y_0$, the set of frequency domain transfer functions $\tilde{c}'_{01},...\tilde{c}'_{0N}$ 
that contain independent noise is the solution to the matrix equation \cite{Allen:1999ct}

\begin{equation}
\begin{bmatrix}
    \tilde{c}'_{01}(f_b) \\
    \tilde{c}'_{02}(f_b) \\
    \vdots \\
    \tilde{c}'_{0N}(f_b)
\end{bmatrix}
=
\begin{bmatrix}
    \tilde{c}_{11}(f_b) &\dots  & \tilde{c}_{N3}(f_b) \\
    \tilde{c}_{12}(f_b) & \dots  & \tilde{c}_{N2}(f_b) \\
    \vdots & \ddots & \vdots \\
    \tilde{c}_{1N}(f_b) & \dots  & \tilde{c}_{NN}(f_b)
\end{bmatrix}^{-1}
\begin{bmatrix}
    \tilde{c}_{01}(f_b) \\
    \tilde{c}_{02}(f_b) \\
    \vdots \\
    \tilde{c}_{0N}(f_b)
\end{bmatrix}
\end{equation}

These independent transfer functions can then be used for noise subtraction as described in Equation \ref{eq:freqsub}. This 
process allows additional sensors that may witness different features of the same 
noise source to be added to the noise subtraction process without risking oversubtraction. 

\subsection{Calculation of Coupled Noise}
Advanced LIGO data is not stationary on the time scale of hours 
\cite{O1DQ, walker:2018}, meaning the transfer functions used to subtract 
each noise source will vary over the time period that the noise subtraction is applied.  
This necessitated the development of methods to understand the 
stability and accuracy of the transfer function estimation on long timescales. 

\begin{figure}[t]
  \centering
  \includegraphics[width=\textwidth]{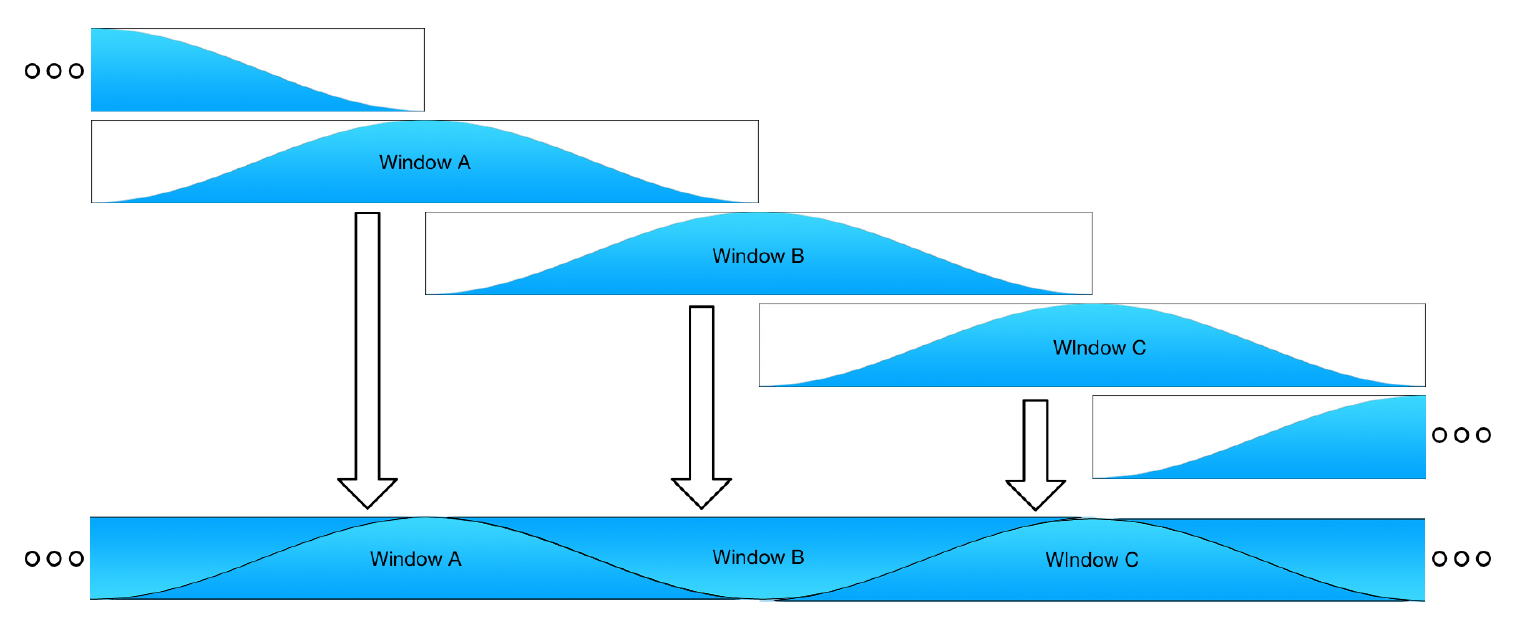}
  \caption{Visualization of how transfer function measurements are tiled in time. %
  Transfer functions are measured in time windows (typically 1024 seconds) with 50\% overlap. For a given %
  time, the transfer function between the witness sensor and $h(t)$ is measured %
  and the witness data are filtered to generate their projected contributions to $h(t)$. %
  A Hann window is applied to each section of projected data before adding them together, %
  resulting in a single projected $h(t)$ time series that has incorporated the time %
  dependence of the transfer functions.}
  \label{fig:methods}
\end{figure}

High amplitude non-Gaussian instrumental artifacts that can impact the measurement of transfer 
functions are removed from the strain data before transfer functions are calculated. This is done by applying an inverse 
Tukey window that zeroes the data containing each instrumental transient. Instrumental transients 
are identified for removal by marking any times where the whitened time series exceeds a value of 100. 
This process is identical 
to the windowing done in \cite{GW170817}. A continuous measurement for long stretches of data 
is approximated by calculating transfer 
functions with overlapping finite measurement windows, called ``sections.'' A visualization of this process 
is shown in Figure \ref{fig:methods}.
After calculating the transfer functions and projected noise contributions for each 
individual window, each section of projected noise is multiplied by a Hann window and 
smoothly added together with 50\% overlapping sections. 

In the case that the transfer function is truly constant, this 
method is identical to applying a single transfer function over the entire period. 
The transfer function for each witness sensor is constructed to be 
uncorrelated with the transfer functions from other witness sensors, resulting in  
noise projection time series that are also independent. This allows each noise time 
series to be subtracted from the strain data independently. Once all targeted noise 
contributions are subtracted, we refer to the data as ``cleaned''. 

\subsection{Workflow Implementation}\label{sec:workflow}

One of the key features of this implementation is the throughput at which the subtraction 
can be done over long stretches of data. The pipeline takes advantage of the 
Pegasus workflow methods implemented in the PyCBC software package 
\cite{deelman2015pegasus,Usman:2015kfa,pycbc-github}, which allows for 
parallelized calculation of transfer functions. Since the transfer functions for different 
noise sources can be calculated independently, the workflow was able to measure 
transfer functions for each noise source and generate projected strain data in parallel. 
In addition, the data set was broken up into distinct sections of continuous detector 
operation that were processed in parallel. 
The limiting factor in the subtraction process  
is the availability of computing nodes. Applying this method using available resources
with 14 witness sensors allowed for two weeks of data from one detector, 
approximately 65 gigabytes, to be processed in only a few hours.  

\section{Noise sources}\label{sec:noise}

During O2, multiple sources of linearly coupled noise were identified. These fell into two 
main categories: beam jitter noise that led to broadband noise contributions and narrow line artifacts 
from power mains and calibration lines. Both of these noise classes were identified and subtracted 
for analyses on previously published events, as described in \cite{Driggers:2017}.
This section describes each noise source and the witness sensors used in the subtraction process. 

\subsection{Jitter Noise}

The main source of linearly coupled noise identified during O2 was related to jitter of the 
pre-stabilized laser (PSL) beam in angle and size \cite{Driggers:2018pe, Kwee:2012, alog:water}. The PSL is responsible 
for generating the frequency- and intensity-stabilized input laser beam that is injected 
into the interferometer. Upgrades to this subsystem undertaken 
in preparation for O2 led to different configurations of the PSL between Hanford and Livingston.

The configuration of the PSL at Hanford during O2 included the addition of a 
high powered oscillator (HPO) that was designed to increase the laser power injected into 
the interferometer up to 200 W \cite{Kwee:2012,Asai:2015aligo}. The optical components used in the HPO required 
continuous heat dissipation via water cooling. Vibrations from water flow coupled to the 
table that supports the optical components used to control the beam angle, introducing 
jitter in beam angle and size \cite{Driggers:2018pe, alog:water}. 

Fluctuations in beam angle are measured using quadrant photodiodes that sense the light 
reflected from the input mode cleaner (IMC) \cite{alog:quadpd}, which is used to filter 
higher order optical modes from the input beam. In February 2017, an additional sensor 
sensitive to radial beam distortions was installed \cite{alog:bullseye}. In total, 7 readouts of beam angle and size (4 derived from quadrant photodiodes and 3 derived from the bullseye photodiode) were used to measure and subtract noise due to beam jitter.

During O2, the coupling of beam jitter into the output of the detector was further complicated 
by the presence of an axially asymmetric point absorber that was present on one of the test masses 
at Hanford \cite{alog:absorber}. Thermal deformations 
are generally corrected with the use of the the Thermal Compensation System (TCS), which heats 
and deforms the mirrors \cite{Brooks:2016}, but this system is not capable of compensating for 
a pointlike deformation. This deformation may have caused beam size and angle fluctuations 
to more strongly couple into the gravitational-wave strain data.
Mitigating beam jitter noise required replacement of the HPO stage in the PSL and the test mass with the point absorber. 
Due to the invasive nature of this work, mitigation was not possible until after the end of the observing run.

Jitter noise related to beam size and beam angle fluctuations was present at Hanford throughout all of O2, with increased coupling towards 
the end of the run. Variations in the beam angle led to broadband noise contributions, while variation 
in beam angle was coupled strongly at mechanical resonances of optic mounts between 
100 and 700 Hz. The sensors used to 
witness these noise sources were digitally sampled at 2048 Hz, which sets the 
maximum frequency at which this jitter noise can be subtracted at 1024 Hz. The broadband 
coupling may have introduced noise above this frequency, but is not addressed in this work. 

At Livingston, the HPO was not included in the O2 configuration, and no asymmetries in the test 
masses were noted, leading to no noticeable jitter coupling in the gravitational-wave strain data. 
For this reason no jitter subtraction was done with the Livingston data, 
which accounts for the lack of broadband noise subtraction seen in the spectrum 
shown in Figure \ref{fig:compare_asd}.  

\subsection{Line Artifacts}\label{sec:lines}

The gravitational-wave strain data demonstrates several noise features that are narrowband, 
appearing as sharp lines in the frequency domain that can affect long duration searches and parameter estimation. 
The strain data contains excess noise at 60 Hz and its harmonic frequencies at both sites due 
to coupling of the power mains. These lines can be subtracted out using 3 witness sensors that directly 
measure the 3-phase voltage provided by the mains power grid at each observatory. 
In addition, the calibration lines discussed in Section 
\ref{calibration} are applied using two methods. One set of calibration lines are digitally 
injected into actuation signals that control the position of the optics. Each digital 
excitation signal can be subtracted using the recorded excitation at the injection point. 
A second set of calibration lines 
are applied to the test masses via radiation pressure using the photon calibrator \cite{Karki:2016} 
and can be measured and subtracted using a single photodetector that monitors the power of the photon 
calibrator beam.

\section{Diagnostics}\label{sec:diagnostics}

\subsection{Sensor Safety}

Before using the witness sensors described in Section \ref{sec:noise} to subtract correlated noise, 
each sensor's sensitivity to gravitational waves, or ``safety'',
was estimated. To establish safety, a series of sine-Gaussian waveforms 
were injected into the detector to excite the degree of freedom that is 
sensitive to gravitational waves \cite{Biwer:2017}. If an excitation of this degree of freedom 
coupled into the readout of any witness sensors in a statistically significant 
way \cite{Smith:2011hv}, those sensors were considered capable of accidentally 
subtracting away real gravitational-wave signals and were marked as unsafe. 
All of the witness sensors used for noise subtraction were determined to be 
incapable of witnessing and subtracting away gravitational-wave signals.

\subsection{Recovery of Simulated Compact Binary Coalescence Signals}\label{sec:cbcsigs}

\begin{figure}[t]
\centering
    \includegraphics[width=0.75\textwidth]{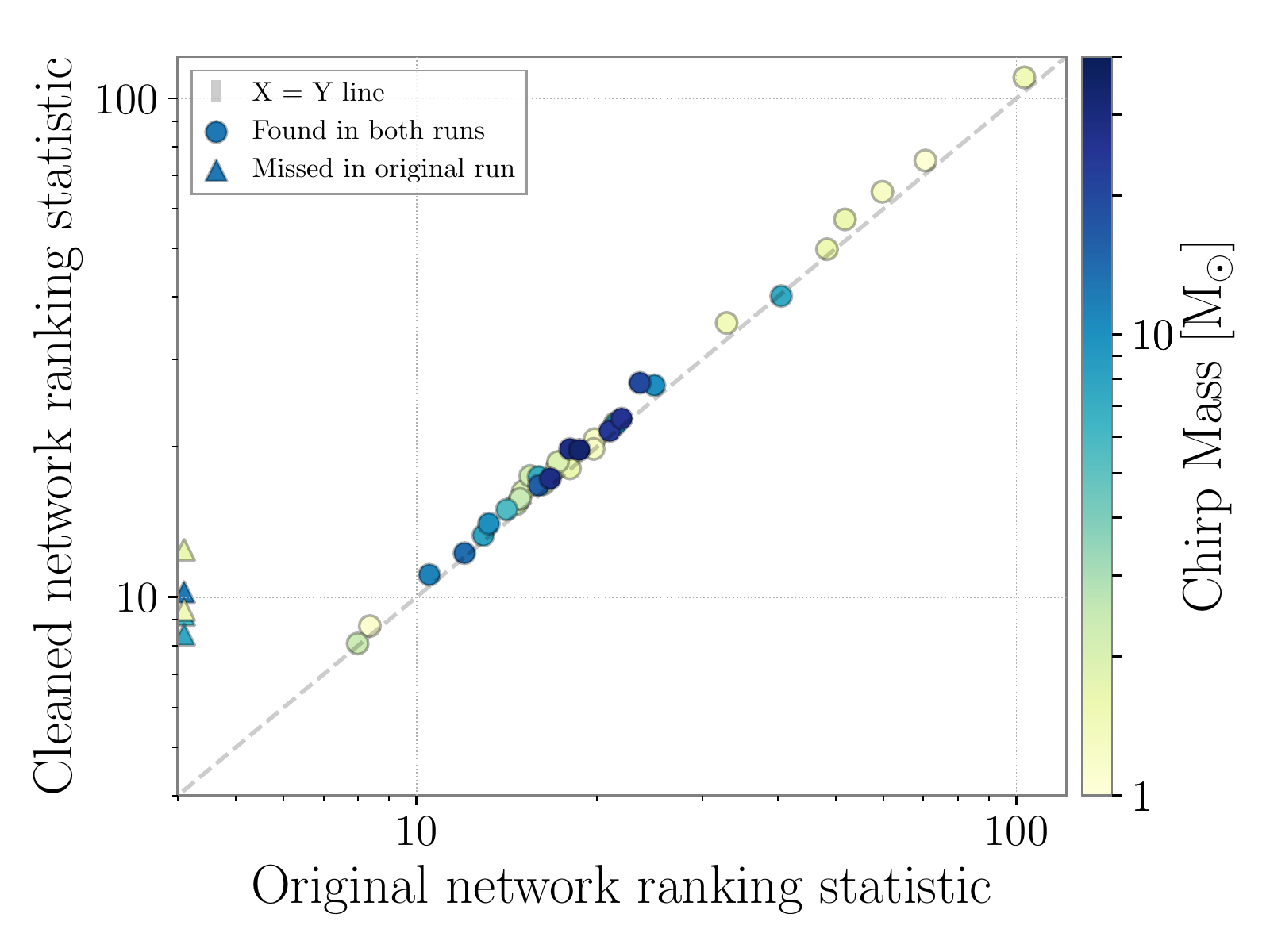}
    \caption{Recovered network ranking statistic for simulated gravitational %
    wave signals before and after applying noise subtraction. %
    The colorbar indicates the chirp mass \cite{Cutler:1994} of each event, which spans a large %
    astrophysical parameter space including binary neutron Star, neutron star - black Hole, and binary black hole signals. %
    After noise subtraction, the simulated signals are recovered with a network %
    ranking statistic that is greater than or equal to the ranking statistic %
    without noise subtraction. In addition, several quiet simulated signals that were below %
    the threshold of the search pipeline were recovered after noise subtraction due to being below the minimum threshold of signal to noise ratio of 5.5 in both detectors. %
    These are indicated with triangles.}
\label{sigs}
\end{figure}

To ensure the noise subtraction process would not corrupt an 
astrophysical signal, a set of simulated compact binary coalescence (CBC) waveforms 
was digitally inserted over five days worth of aLIGO data from both detectors. 
The data set containing these simulated signals was processed 
by the PyCBC astrophysical search algorithm \cite{Usman:2015kfa,pycbc-github},
used to search for signals from stellar-mass neutron star and black hole binaries,
in order to compare the recovery of the simulated signals before and after noise subtraction. 
For each recovered signal, a coincident ranking 
statistic that represents the significance of an event found in multiple detectors in the 
detector network is calculated. Figure \ref{sigs} shows the 
recovered coincident ranking statistic of each simulated signal before and after subtracting 
noise from the data set. After noise subtraction, all of the simulated signals were 
recovered with a ranking statistic that is consistent with or better than the  
ranking statistic before subtraction. In addition, there is a population of 
simulated signals that were not recovered in the original analysis but were 
found as coincident events after noise subtraction.
As a final test, hardware injected CBC signals \cite{Biwer:2017} were  
successfully recovered after performing noise subtraction. 
These hardware injections were recovered with increases in ranking statistic consistent 
with changes seen in software injections. 

\subsection{Simulated Noise Tests}

To verify that the noise subtraction process is effective for generic noise 
sources, artificial noise was added to aLIGO strain data and processed using the same method. 
The first test attempted to subtract artificial correlated noise. 
This noise was constructed by generating 
Gaussian noise, passing it through a transfer function that had similar features to the jitter transfer function, 
and summing it into the strain data. When provided with the strain data and the 
Gaussian noise, the noise subtraction algorithm was able to 
reconstruct the transfer function used to project the Gaussian noise into the 
strain data and subtract out the excess noise. The amplitude spectral density of 
the resulting data was consistent with the original data to within $\pm3\%$ 
at all frequencies.

The second test was to subtract out random, uncorrelated noise which had 
\textit{not} been added to the strain data. 
When provided with the strain data and the uncorrelated Gaussian noise, 
the algorithm subtracted a minimal amount of random noise. 
Similarly, the amplitude spectral density of the 
resulting data was consistent with the original 
data to within $\pm2\%$ at all frequencies. 

\subsection{Effect on Calibration}\label{calibration}

One important feature of aLIGO data is the presence of continuous, narrowband 
sinusoidal injections, or ``calibration lines'', 
which are used to calibrate the data \cite{Viets:2018cal}. 
This calibration is performed on data that does not have noise subtracted, therefore tests were 
conducted to ensure that the calibration of the data was still valid after cleaning. 
A set of noise subtracted data was produced using data from Hanford that did not subtract away 
calibration lines in order to measure the impact of broadband noise subtraction on the 
data calibration process. Both the cleaned and uncleaned strain data were demodulated at 
the calibration line frequencies and the amplitude and phase were averaged in 300 second bins. 
The amplitude ratio and phase offset of each resulting measurement were calculated and are 
used as metrics for consistency. 

To accumulate a statistically significant measurement of the 
calibration line consistency, 6.65 days of data were analyzed and the  
$1~\sigma$ errors on the distribution of amplitude ratios and phase offsets are reported. 
For the 36.7 Hz and 
1083.7 Hz calibration lines at Hanford, the amplitude ratio was consistent with 1 to within $\pm0.014\%$ and 
the phase offset was consistent with $0^{\circ}$ to within $\pm0.0078^{\circ}$. 
The 331 Hz calibration line (Located at a frequency where 
a non-negligible amount of power is expected to be subtracted off due to beam jitter) has 
an amplitude ratio that is consistent with 1 to within $\pm0.15\%$ and a phase offset that is 
consistent with $0^{\circ}$ to within $\pm0.087^{\circ}$. As typical calibration uncertainties are $\pm 4 \%$ \cite{Cahillane:2017cal}, these measurements confirm that the noise 
subtraction process did not significantly impact the overall calibration of the strain data.

\subsection{Impact of Nonstationary Data}\label{sec:nonstat}

While generally stable, the witness sensors used for transfer function estimation 
sometimes contain transient noise. In cases where the witness sensor 
has transient excess power that is not linearly correlated to the gravitational-wave strain, the 
transfer function is overestimated and the noise subtraction algorithm removes too much 
projected noise from the strain data. 
However, when the transient noise is linearly correlated to 
the gravitational-wave strain, transient noise can be subtracted from the 
gravitational-wave strain data. This linear subtraction of transient noise was commonly found during 
periods of transient noise in the power mains. 

The most impactful cases of oversubtraction due to excess power that is not linearly correlated with 
the gravitational-wave strain were noticed during review of the cleaned data set, and
occur when there is transient noise in the photon calibrator used to inject 
calibration lines into the detector. To avoid this overestimation, the noise subtraction 
process is halted for 3 seconds around these transient noise artifacts. 
As these noise artifacts last less than one second, this veto period was chosen 
to ensure that the effect of the transient on transfer function measurement was completely mitigated. 
Once times where witness sensors contain transient noise are removed, the nearby 
noise subtracted data shows no evidence of oversubtraction as compared to time periods disjoint
from the excess noise. 

Additional oversubtraction may occur if a feature of a witness sensor is spuriously correlated with 
the gravitational wave data. While such features are not observed on the timescales that the 
subtraction process is computed over, narrowband noise features from beat notes in the 
photon calibrator system may appear when signals are averaged on the timescale of multiple hours. 
The total bandwidth affected by these spurious correlations is less than 0.1 Hz and can be 
removed from long timescale analyses with the use of notch filters \cite{Covas:2018}.

\section{Results}\label{sec:results}

\subsection{The O2 Data Set}

The noise subtraction algorithm was used to clean the entire data set from 
Advanced LIGO's second observing run, which spanned 9 months. The final version of calibrated data \cite{Viets:2018cal,Cahillane:2017cal} 
was used as the input to 
the noise subtraction pipeline. For computational efficiency, data which were 
considered unfit for astrophysical analysis \cite{O1DQ,Nuttall:2015dqa,Nuttall:2018dq,Berger:2018dq} were not processed. 
Additional time losses were due to removal of time periods corrupted by bandpass filters applied in the subtraction process and the excision of data where witness sensors 
were unsuitable for reliable transfer function measurement, as noted in Section \ref{sec:nonstat}.
In all, only 
0.05\% of 
strain data was discarded as a result of the noise subtraction process. The final 
cleaned data set contains 118 days of coincident data. This value 
is greater than the coincident livetime reported in \cite{GW170817} due to the inclusion 
of additional time with updated calibration \cite{Cahillane:2017cal,Viets:2018cal}. 

\subsection{Effects on the Noise Curve}

\begin{figure*}[t]
\centering
    \includegraphics[width=\textwidth]{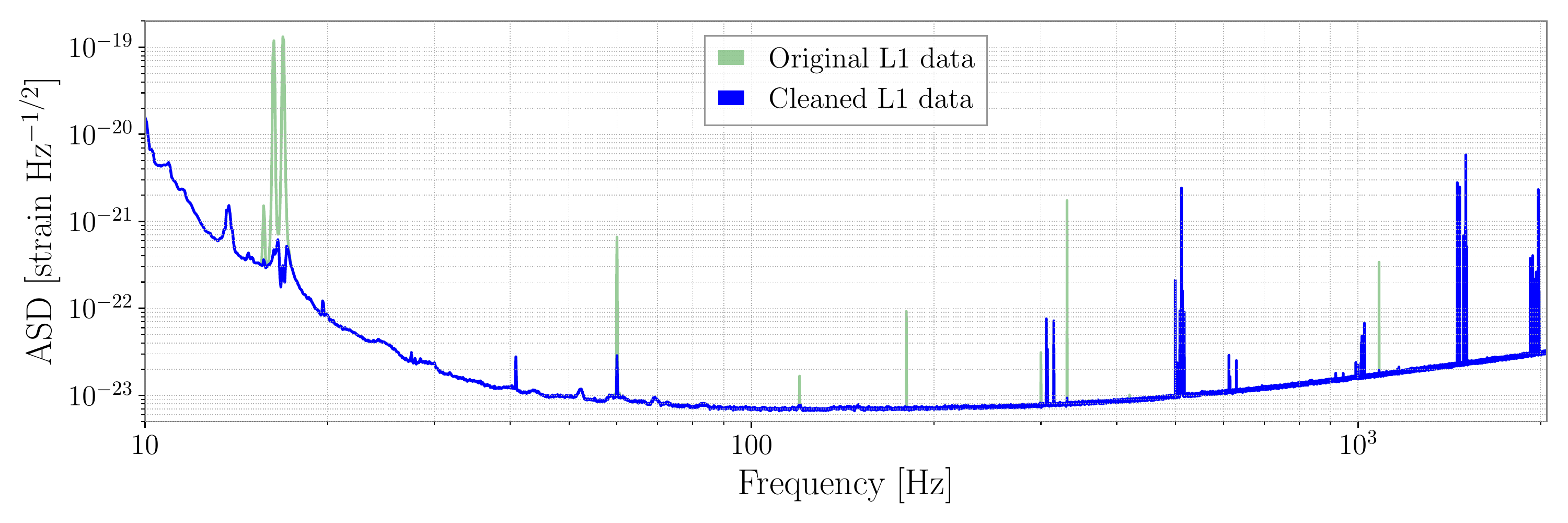}
    \includegraphics[width=\textwidth]{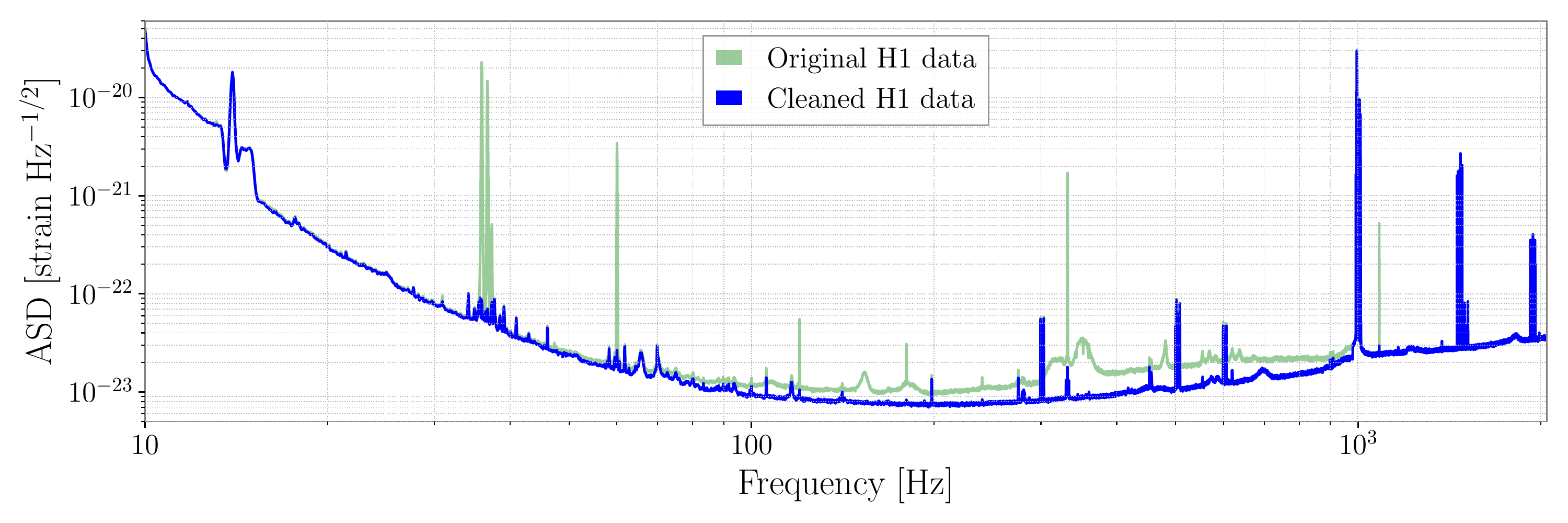}
    \caption{Top: Amplitude spectral density (ASD) of Livingston (L1) gravitational wave strain data before %
    (green) %
    and after (blue) noise subtraction from a representative day of data during O2. Narrowband features from calibration lines %
    and power mains %
    were subtracted from the L1 strain data. There were no broadband noise sources with an 
    appropriate witness sensor that could be subtracted from the L1 strain data. %
    Bottom: Amplitude spectral density of Hanford (H1) gravitational wave strain data before and %
    after noise subtraction from a representative day of data during O2. In addition to narrowband features from calibration lines %
    and power mains, broadband noise was subtracted between 80 - 1000 Hz using %
    beam jitter witness sensors.}
\label{fig:compare_asd}
\end{figure*}

The noise subtraction process is capable of removing both narrowband and broadband 
spectral features. 
Figure \ref{fig:compare_asd} shows the amplitude spectral density of the strain data from the 
Hanford and Livingston detectors before and after noise subtraction. 
The narrow lines removed at 33, 60, 120, 180, 331, and 1083 Hz, 
detailed in Section \ref{sec:lines}, 
are related to detector calibration lines and power mains harmonics.
Broadband subtraction in Hanford data is a result of removing noise related to beam 
jitter. 
Due to the 2048 Hz sampling rate of the witness sensors and a low pass 
filter applied to reduce corruption near the Nyquist frequency, broadband 
noise is only subtracted up to 1024 Hz. A high pass filter applied at 13 Hz 
set the minimum frequency at which broadband noise was subtracted. 

The same procedure was used to address noise sources present in the Livingston 
detector. Line artifacts due to calibration lines and harmonics of the power 
mains were removed. 
As previously noted, beam jitter noise did not contribute significantly 
to the Livingston data and was not subtracted.

\begin{figure}[t]
\centering
    \includegraphics[width=\textwidth]{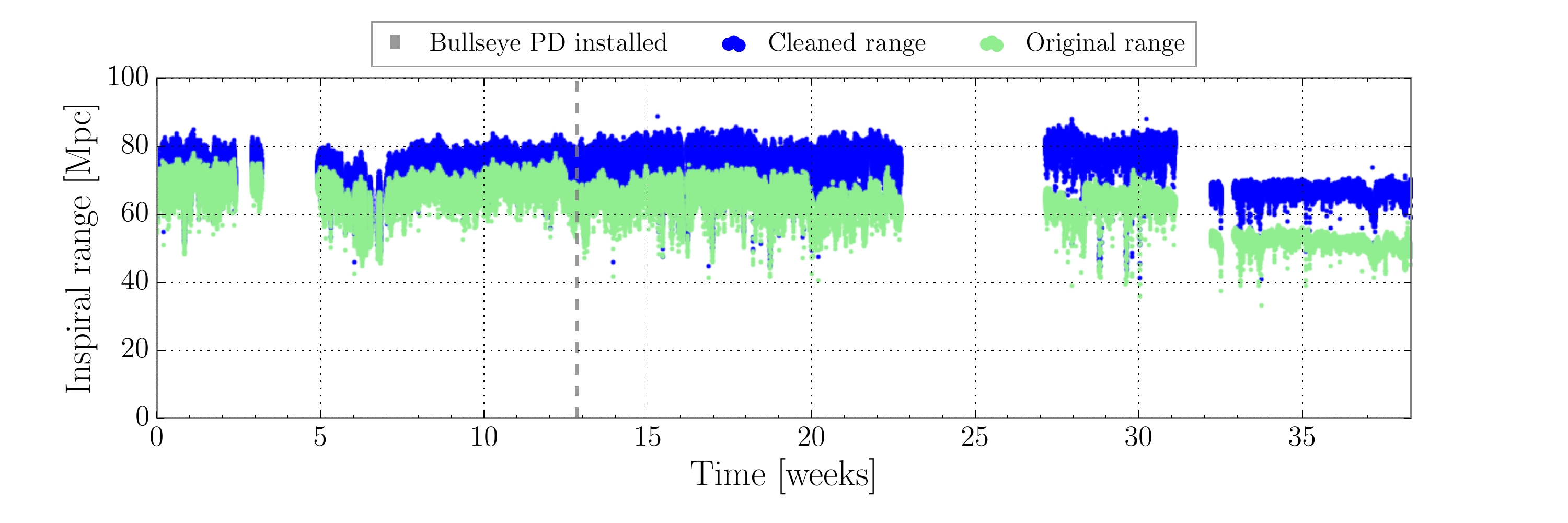}
    \caption{Inspiral range of the Hanford (H1) detector over the course of O2 before (green) %
    and after (blue) noise subtraction. The dashed line indicates the installation %
    of the bullseye photodiode, a witness sensor used for subtraction of noise due %
    to beam jitter. %
    The large decrease in range after week 32 for both the Original and Cleaned range was %
    due to the impact of an earthquake near the site \cite{alog:moneq}. %
    Livingston (L1) had no broadband noise subtraction, the increase in inspiral range %
    was negligible and is not shown.}
\label{fig:O2range}
\end{figure}

We can characterize the benefit of the noise subtraction process with 
the ``inspiral range'', which is the average distance at which a detector could 
observe a BNS system (1.4 - 1.4 $\mathrm{M_\odot}$) at a signal to noise ratio (SNR) of 8. The inspiral range during O2 at Hanford before and after noise subtraction is shown in Figure \ref{fig:O2range}.
After noise subtraction, the inspiral range at Hanford increased by $\sim20\%$ when averaged 
over all of O2 with a peak increase of $\sim50\%$ towards the end of the observing run. 
The change in inspiral range at Livingston was negligible 
over the course of O2 due to the lack of broadband noise subtraction. 
The effect of noise subtraction on the overall network sensitivity of the detectors 
is discussed in Section \ref{sec:sensitivity}.

\subsection{Effect on Astrophysical Analyses}\label{sec:sensitivity}

\begin{figure}[t]
\centering
    \includegraphics[width=\textwidth]{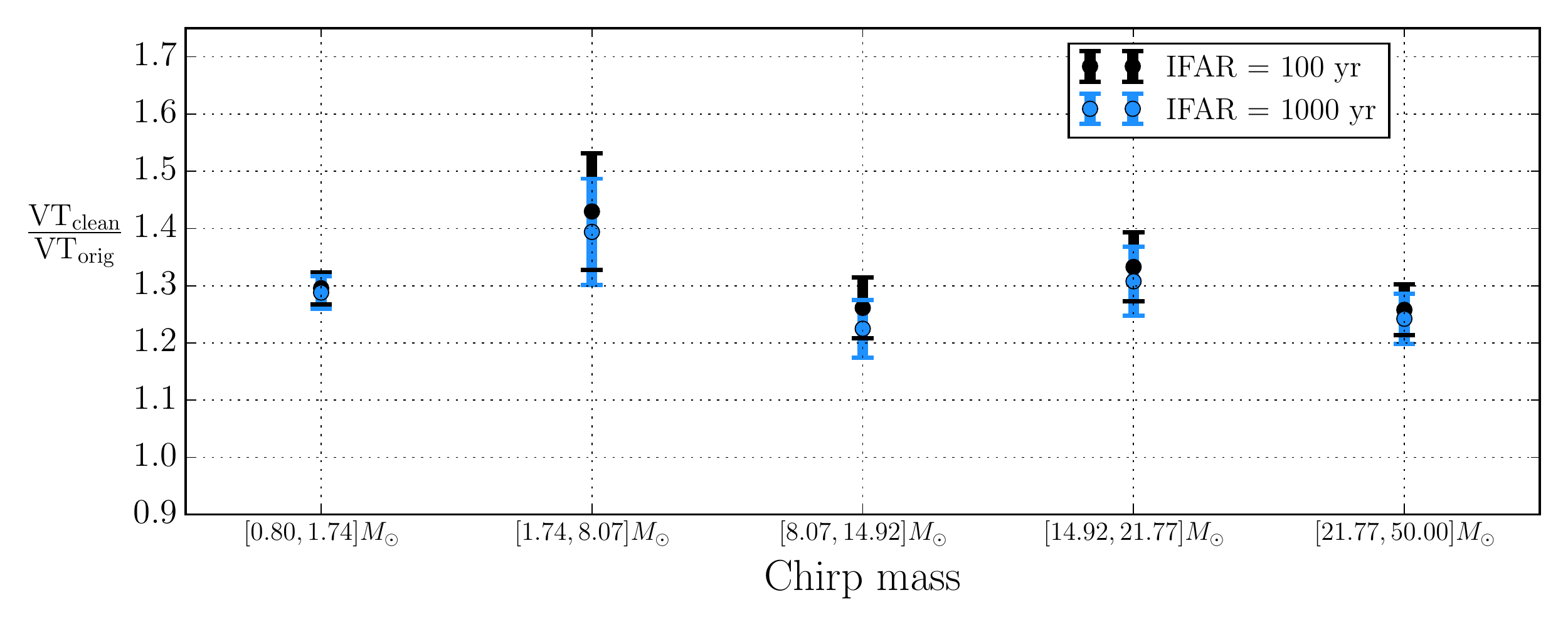}
    \caption{The ratio of volume-time (V-T) the PyCBC search was sensitive to during O2 %
    before (original) and after noise subtraction (clean) binned by chirp mass. %
    Black represents the volume-time that the search was sensitive to for signals with %
    an inverse false alarm rate (IFAR) of 100 years, while Blue corresponds to signals %
    with an IFAR of 1000 years. Error bars show 1 sigma error. On average, a 30\% %
    increase in V-T was measured over the course of O2.}
    \label{CBC-VT}
\end{figure}

The figure of merit used for quantifying sensitivity of a search compact binary coalescences is 
the sensitive volume of the search multiplied by the time duration of analyzed data, which is known 
as volume-time (V-T). While this volume can be approximated using the inspiral range as a measure of 
sensitive distance, that method does not fully account for the effects of data containing non-Gaussian noise artifacts on 
astrophysical search sensitivity, as well as the sensitivity of the entire interferometer network. 
V-T can be measured by injecting n population of simulated gravitational-wave signals into the 
data and attempting to recover them with a search pipeline \cite{Usman:2015kfa}. 
For each search pipeline, a background distribution is generated that excludes coincident events 
in order to estimate the effects of detector noise on the search algorithm. 
Each recovered signal is then compared to this background and assigned an inverse false alarm 
rate (IFAR) that quantifies how likely it is that such a signal was caused by coincident 
instrumental artifacts rather than an astrophysical source. 
To estimate the increase in sensitivity due to noise subtraction, V-T of the PyCBC search was measured 
before and after noise subtraction using identical injection sets. As multiple values can be used as a 
cutoff to determine if a signal is recovered, we examined the V-T for IFAR values of both 100 years and 
1000 years. 
The ratio of V-T before and after noise subtraction binned by chirp mass 
is shown in Figure \ref{CBC-VT}. 

The large increase to the sensitive volume of the detector network, combined with a negligible 
reduction in available coincident time, led to a significant increase in V-T over the course of O2. 
Averaging over all mass bins, a 30\% increase in V-T was measured over the course of O2. 
Particularly, the largest gains in sensitivity were realized for chirp mass between 1.74 $\mathrm{M_\odot}$ and 8.07 $\mathrm{M_\odot}$. This is a parameter space that aLIGO 
has not previously detected signals in, and hence has a largely unconstrained rate, 
in addition to being the location of the observed NS-BH mass gap \cite{Littenberg:2015mg,Farr:2011bh,Ozel:2010bh}. 

One observed effect that led to a difference in measured V-T versus V-T extrapolated 
from estimating sensitive volume as a sphere with radius equal to the inspiral range 
was the impact of the noise subtraction process on instrumental artifacts. 
While the noise subtraction process reduced the broadband noise in the detector, 
it did not affect the amplitude of noise artifacts unrelated to the noise sources addressed 
by the noise subtraction pipeline. 
With a lower noise floor and no change in their absolute amplitude, artifacts already present in 
the data were found to increase in SNR. 
As one of the primary limitations of an astrophysical search's ability to recover signals is the 
rate of loud noise artifacts \cite{O1DQ}, these SNR increases limit the increase in V-T due to 
noise subtraction.

\section{Future prospects}
This paper presents an adaptable, parallelized method for subtracting linearly coupled 
noise from aLIGO data. The versatility of this method to remove known sources of noise makes 
it a useful tool to address a variety of noise sources in future observing runs. 
While beam jitter and line artifacts are the only noise sources subtracted from this data set, 
noise from feedback loops used to sense and control the length and alignment of optical 
cavities in the aLIGO detectors have been shown to contribute noise at 
lower frequencies \cite{Driggers:2017} and are potential candidates for subtraction 
in future observing runs. Before the aLIGO's third observing run, the test mass with the point 
absorber at the Hanford detector will be replaced and the PSL configuration will be updated, 
which are expected to reduce noise contributions due to beam jitter. 

The sensitivity gains demonstrated in this paper show that a robust offline noise subtraction 
pipeline is an integral aspect of achieving maximum sensitivity in gravitational-wave detectors. 
The 30\% increase in sensitivity of aLIGO to compact binary coalescences after noise subtraction 
will allow for an increased volume of spacetime to be searched for gravitational waves. 
Although not quantified in this work, the noise subtraction process will also lead to general 
increases in the sensitivity of searches for gravitational waves using aLIGO data, such as 
those for continuous waves \cite{Riles:2017}, stochastic \cite{Regimbau:2011}, 
and unmodeled burst sources \cite{Lynch:2015yin,Klimenko:2008fu,Cornish:2014kda}. 

\section{Acknowledgments}

We would like to thank the aLIGO commissioners and the Detector Characterization group for 
identifying and characterizing the noise sources addressed,
as well as the PyCBC search group for devselopment of the
injection sets used in this work. We also thank review 
team members Gabriele Vajente and Francesco Salemi for helpful discussions during development, 
along with Jess McIver and Marissa Walker for their comments during the during the internal 
review process. Computing support for this project was provided by the LDAS computing cluster 
at the California Institute of Technology. 
DD acknowledges support from NSF award PHY-1607169.  
LKN received funding from the European Union Horizon 2020 research and
innovation programme under the Marie Sklodowska-Curie grant agreement No 663830. 
TJM acknowledges support from the LIGO Laboratory.
LIGO was constructed by the California Institute of Technology and Massachusetts 
Institute of Technology with funding from the National Science Foundation, and 
operates under cooperative agreement PHY-0757058. This paper carries LIGO Document 
Number LIGO-P1800169.

The authors thank the LIGO Scientific Collaboration for access to the data and gratefully 
acknowledge the support of the United States National Science Foundation (NSF) for the 
construction and operation of the LIGO Laboratory and Advanced LIGO as well as the Science 
and Technology Facilities Council (STFC) of the United Kingdom, and the Max-Planck-Society 
(MPS) for support of the construction of Advanced LIGO. Additional support for Advanced 
LIGO was provided by the Australian Research Council.

\section{References}
\bibliographystyle{unsrt}

\end{document}